\documentclass[a4paper,fleqn,usenatbib]{mnras}

\usepackage{newtxtext,newtxmath}

\usepackage[T1]{fontenc}
\usepackage{ae,aecompl}

\usepackage{graphicx}	
\usepackage{amsmath}	
\usepackage{amssymb}	

\title[Hyperons and quarks in proto-neutron stars]{Hyperons and quarks in proto-neutron stars}

\author[J. Roark et al.]{
J. Roark,$^{1}$
X. Du,$^{2}$
C. Constantinou$^{1}$
V. Dexheimer,$^{1}$
A. W. Steiner$^{2,3}$
and J. R. Stone$^{2,4}$
\\
$^{1}$Department of Physics, Kent State University, Kent, OH 44243 USA\\
$^{2}$Department of Physics and Astronomy, University of Tennessee, Knoxville, TN 37996, USA\\
$^{3}$Physics Division, Oak Ridge National Laboratory, Oak Ridge, TN 37831, USA\\
$^{4}$Department of Physics (Astro), University of Oxford, Keble Road OX1 3RH, Oxford, UK}

\date{Accepted XXX. Received YYY; in original form ZZZ}

\pubyear{2019}

\begin{document}
\label{firstpage}
\pagerange{\pageref{firstpage}--\pageref{lastpage}}
\maketitle

\begin{abstract}

In this work, we study matter in the cores of proto-neutron stars,
focusing on the impact of their composition on the stellar structure.
We begin by examining the effects of finite temperature (through a
fixed entropy per baryon) and lepton fraction on purely nucleonic
matter by making use of the DSH model . We then turn our
attention to a relativistic mean-field model containing exotic degrees
of freedom, the Chiral Mean Field (CMF) model, again, under the
conditions of finite temperature and trapped neutrinos. In the latter,
since both hyperons and quarks are found in the cores of large-mass
stars, their interplay and the possibility of mixtures of phases is
taken into account and analyzed. Finally, we discuss how stellar
rotation can affect our results.
 
\end{abstract}




\section{Introduction}

Neutron star temperatures become much smaller than the neutron Fermi energy within a few minutes after a neutron star is born. This is not the case for so-called proto-neutron stars (PNS) or for the remnant objects of binary NS mergers. In the former instance, the core temperature can reach a few tens of MeV \cite{Burrows:1986me,Pons:1998mm} whereas, in the latter, it can reach up to $100$ MeV in certain regions of hypermassive NS's \cite{Galeazzi:2013mia,Most:2018eaw}. At $T=0$,  hyperons, being more massive than nucleons, appear in stars at the usual threshold of a couple of times the equilibrium density of symmetric nuclear matter, $n_s = 0.16 \pm 0.01$ fm$^{-3}$, whereas the quark phase is reached when the strong coupling becomes weak enough, at a few times $n_s$. At high temperatures, the chemical potentials (at a given baryon density) of the constituent particles decrease at varying degrees depending on their effective masses (consequences themselves of interactions with the surrounding medium) with corresponding changes in their populations.

Immediately after  a PNS has been born, matter in its core is opaque to neutrinos and only after a time-scale of $10 -15$ s \cite{Burrows:1986me,Burrows:1991,keil95} does it start to cool, via URCA type processes, when it becomes transparent to neutrinos. This evolution depends on the distribution of baryonic/quark and leptonic constituents in the stellar interior and, by extension, on the equation of state (EoS) of hot and dense matter. Several works have studied the effects of trapped neutrinos and temperature in stars with hyperons \cite{Prakash:1997,Dexheimer:2008ax,Yasutake:2012dw,Masuda:2015kha,Lenka:2018ehb,balberg97,vidana03,mornas05,marques17}, chiral partners \cite{Dexheimer:2012eu} and stars with a deconfinement phase transition \cite{Prakash:1997,Steiner:2000bi,Pons01eo,Menezes:2003pa,SchaffnerBielich:2007mr,Nakazato:2008su,Gu:2008zze,Bombaci:2009jt,Lugones:2009ms,Lugones:2010gj,Shao:2011nu,Tatsumi:2011tt,Bombaci:2011mx,
Yasutake:2012dw,Dexheimer:2012eu,Carmo:2013oba,Hempel:2013tfa,Hempel:2015vlg,Masuda:2015kha,Olson:2016cxd,Mariani:2016pcx,Bombaci:2016xuj,Marquez:2017bqm,lugones98,carter00,steiner01,
menezes03,nicotra06,sandin07,glendenning95}.

In this work, we perform a thorough investigation of the composition and structure of PNS's modeled by fixed lepton fraction while drawing comparisons with deleptonized (with respect to neutrinos) $\beta$-equilibrated NS's in the context of several models. These models possess different degrees of freedom together with different interactions. An overview of these models is given in section II with results presented in section III, where we also briefly discuss how the stellar particle population is modified by stellar rotation. Our conclusions are given in section IV.

\section{Formalism}

We begin by presenting different models that fulfill standard nuclear and astrophysical constraints.

\subsection{DSH model}
Our prototype model for matter with only nucleonic degrees of freedom
is the recent set of EoS's constructed in Ref.~\cite{Du18ha},
hereafter ``DSH''. These EoS's were built from a parametrized
phenomenological model for homogeneous nucleonic matter that can be
used over a wide range of densities, temperatures, and electron
fractions. It is designed to match (i) the virial expansion at low
densities, (ii) nuclear structure experiments which probe nearly
isospin-symmetric matter near $n_s$, (iii) chiral effective theory
which provides a description of pure neutron matter, and (iv) NS mass
and radius observations. The phenomenological
  parameterization contains several parameters which are selected
  randomly and, then, models which are physically disallowed (for
  example, because their maximum mass is not sufficiently large or
  because the nucleon effective mass becomes negative) are
  removed.
  
The DSH formalism provides an infinite family of EoS's selected
according to a probability density to match these four sets of
constraints. The range of neutron star radii is somewhat
  small, even though Ref.~\cite{Du18ha} takes care to include the
  relevant uncertainties. Three effects give rise to this result: (i)
  DSH presumes no strong phase transitions in dense matter, (ii) the
  observational requirement that the maximum mass must be larger than
  about $2~\mathrm{M}_{\odot}$, and (iii) the push towards smaller
  radii, as suggested by neutron star mass and radius observations. In
this work, we randomly select 10 DSH EoS parameterizations to give the
reader an idea of the remaining uncertainties in the quantities of
interest and at the same time a benchmark for understanding the
effects of temperature and neutrinos in a model that does not include
exotic degrees of freedom.
  
\subsection{CMF model}
The Chiral Mean Field (CMF) model is based on a non-linear realization of the SU(3) sigma model. It is an effective quantum relativistic model that describes hadrons and quarks interacting via meson exchange.  This formalism is based on chiral invariance, meaning that the particle masses originate from interactions with the medium and, therefore, decrease at high densities \cite{Dexheimer:2009hi,Hempel:2013tfa}. The deconfinement of quarks is mimicked by the introduction of an order parameter $\Phi$. The hadronic coupling constants of the model are calibrated to reproduce the vacuum masses of baryons and mesons, nuclear constraints for isospin-symmetric and asymmetric matter at saturation, and reasonable values for the hyperon potentials. The quark coupling constants are constrained using lattice QCD data, as well as information about the QCD phase diagram for isospin-symmetric and asymmetric matter. As a consequence, this formalism reproduces the nuclear liquid-gas phase transition, as well as the deconfinement/chiral symmetry restoration phase transitions expected to be found in the QCD phase diagram with critical points and crossover regions. Finally, the model is successful in reproducing perturbative QCD (PQCD) results for both NS and PNS matter at high densities \cite{Roark:2018uls,Kurkela:2016was}.

Since the CMF model includes deconfinement to quark matter, it is important to note that, as explained in detail in Ref.~\cite{Hempel:2013tfa} and references therein, whenever two or more quantities (including baryon number) are conserved globally in coexisting macroscopic phases with different compositions, non-congruent phase transitions are present. This is usually referred to as Gibbs construction when modeling NS's with globally imposed charge neutrality (NS GCN). Alternatively, if the surface tension of quark matter is too high, local electric charge neutrality is established between the phases instead. This is usually referred to as Maxwell construction when modeling NS's and denoted here as (NS LCN). A discussion about how surface tension values can depend on density, temperature, neutrino trapping, and magnetic fields can be found in Ref.~\cite{Lugones:2018qgu}.

As already discussed, PNS matter is modeled by imposing lepton fraction conservation. Recently, Ref.~\cite{Roark:2018uls} derived in detail the formalism necessary to describe non-congruent phase transitions in the case of PNS matter. Lepton fraction is conserved globally, since there is no long-range force (equivalent to Coulomb) associated with it \cite{Hempel:2009vp}. Here, for simplicity, we only discuss the non-congruent case for PNS matter where lepton fraction is globally conserved but electric charge is locally conserved in each phase (PNS LCN G$Y_l$). In contrast to our previous work, we now calculate macroscopic stellar structures. In addition, we study a non-physical forced-congruent case for comparison, in which both electric charge and lepton fraction are locally conserved in each phase (PNS LCN L$Y_l$), again, calculating macroscopic stellar structures.  Also for comparison, protoneutron stars featuring only hadrons (PNS H), as well as  neutron stars featuring only hadrons (NS H) will be studied in the following section. 

\begin{figure}
  \includegraphics[trim={1cm 0 0 2.1cm},width=9.5cm]{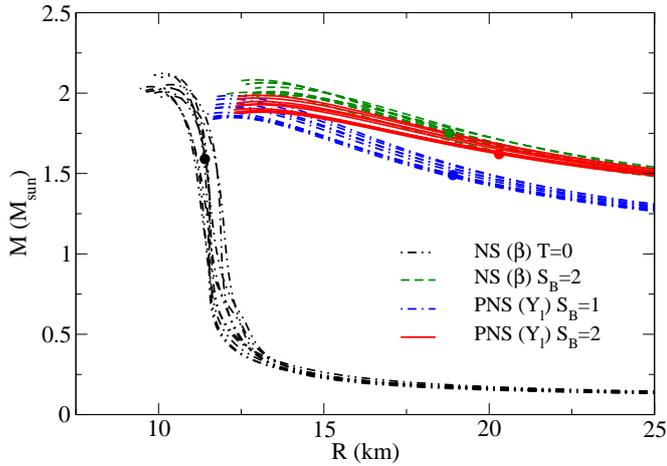}
  \caption{ Mass-radius diagram for 10 different DSH EoS parameterizations  for two $\beta-$equilibrated neutron star and two fixed lepton-fraction proto-neutron star configurations, each at the indicated temperature or entropy per baryon. The circles represent sample stars with the same central baryon number density of $0.6$ fm$^{-3}$.}
  \label{dsh_mvsr}
\end{figure}

\section{Results}

In all cases involving NS matter, calculations are done by imposing charge neutrality and $\beta-$(chemical) equilibrium. The temperature is set to zero, except when otherwise stated. When dealing with PNS matter, trapped neutrinos are included through a fixed (electron and electron neutrino) lepton fraction $Y_l=L/B=0.4$, where $L$ and $B$ are the numbers of  electron-type leptons and baryons, respectively. This typical value of 0.4 comes from numerical simulations of proto-neutron-star evolution 
\cite{Burrows:1986me,Pons:1998mm,Fischer:2009af,Huedepohl:2009wh,peres13}. Finite temperature is included by means of a fixed entropy density per baryon number density $S_B = {s}/{n_B}$ (usually referred to as entropy per baryon). When the entropy per baryon is fixed, it allows the temperature in stars to increase toward the center: For small entropies $S_B \leq 2$, ~$S_B \propto T/n^{2/3}$ (ignoring the week density dependence of the Landau effective mass at high density). Thus, for constant $S_B$, $T$ increases with $n_B$,  becoming larger towards the center of a star. In this section, entropy per baryon is fixed at $S_B =1$ or $2$.  Note that some finite temperature NS and zero temperature PNS results are shown for the sake of comparison.

\begin{figure}
  \includegraphics[trim={1.5cm 0 0 2.8cm},width=9.3cm]{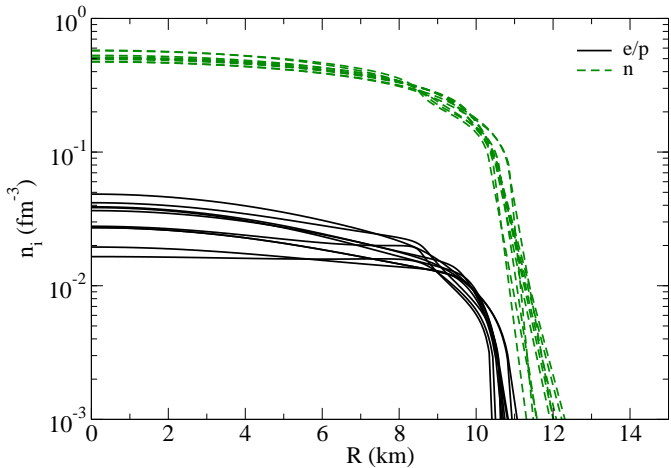}
  \caption{Particle number densities as a function of stellar radius (of a $1.4$ M$_\odot$ star) for cold neutron stars showing 10 different DSH EoS parameterizations.}
  \label{dsh_NS}
\end{figure}

We begin by discussing the influence of lepton fraction and fixed entropy per baryon on the DSH model. Fig.~\ref{dsh_mvsr} shows mass-radius curves obtained by using 10 different DSH EoS parameterizations in the Tolman-Oppenheimer-Volkoff (TOV) equations for each cold or hot configuration (without or with neutrinos). For cold NS matter, a cold neutron-star crust was added to the EoS including an inner crust, an outer crust and an atmosphere \cite{Baym:1971pw}. For the other cases, a hot PNS crust with entropy per baryon $S_B=4$ was applied \cite{Lattimer:1991nc}. The four colors represent NS and PNS evolution snapshots from Ref.~\cite{Pons:1998mm}. On average, neutrino-free $\beta-$equilibrated stars exhibit a small decrease in maximum stellar mass with increasing entropy per baryon (from dashed double-dotted black to dashed green lines) whereas the opposite is true for stars with fixed lepton fraction (from dashed-dotted blue to full red lines). The latter also support less gravitational mass than their $\beta-$equilibrated counterparts at the same entropy per baryon (from red to green lines), due to their higher content of isospin-symmetric matter, which corresponds to a softer EoS.  Curves of constant central density $n_c$ in Fig.~1 are directed diagonally across the plot (for example connecting the circles that represent sample stars with central baryon number density of $0.6$ fm$^{-3}$). That is, at the same $n_c$, increasing $S_B$ and decreasing $Y_l$ leads to more massive stars. Note that, in order to follow the temporal evolution of isolated stars, we would have to follow particular paths of fixed baryon masses, which are not shown here. We revisit this point in our discussion of Fig. \ref{densityVSomega}.

\begin{figure}
  \includegraphics[trim={1cm 0 0 2.81cm},width=9.5cm]{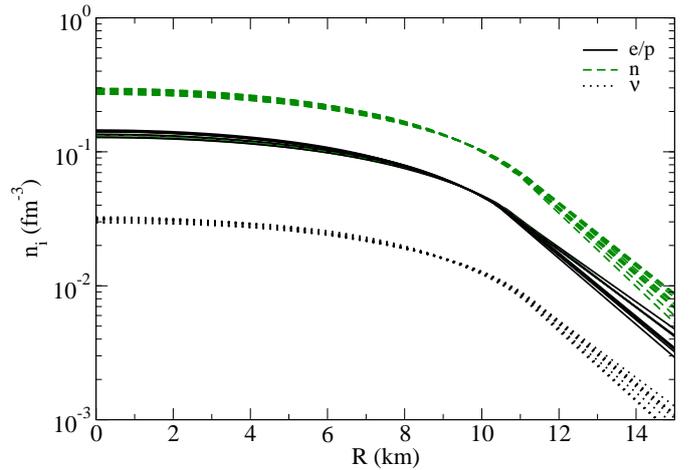}
  \caption{Particle number densities as a function of stellar radius (of a $1.4$ M$_\odot$ star) for proto-neutron stars with fixed entropy per particle $S_B = 2$ showing 10 different DSH EoS parameterizations.}
  \label{dsh_PNS}
\end{figure}

The effects of fixed lepton fraction (together with fixed entropy per baryon) on a $1.4$ M$_\odot$ star can be better seen in Figs.~\ref{dsh_NS} and \ref{dsh_PNS}, which show the population for NS and PNS matter as a function of stellar radius. In the PNS case, the relative amount of protons (with respect to the total baryon number) is much larger than in the NS case. In both figures the amount of electrons equals the amount of protons in order to satisfy charge neutrality. For PNS matter, the amount of neutrinos does not go over $0.04$ fm$^{-3}$.

\begin{figure}
\vspace{3mm}
\includegraphics[trim={1cm 0 0 2.1cm},width=9.5cm]{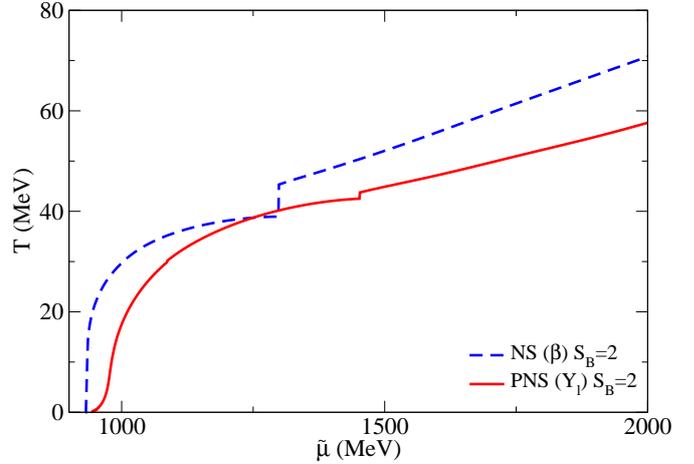}
\caption{\label{Tvsmu_Bhat} Temperature vs (modified) chemical potential phase diagram for neutron-star matter with locally conserved electric charge and proto-neutron-star matter with locally conserved electric charge and lepton fraction, both calculated at a fixed entropy per baryon in the CMF model.}
\end{figure}

To study the effects caused by the presence of exotic matter in proto-neutron stars and the interplay of the appearance of hyperons and quarks, we now make use of the CMF model. Once more, we fix the entropy per baryon and allow the temperature to change inside each star. The resulting temperature profile can be seen in Fig.~\ref{Tvsmu_Bhat} for the PNS LCN L$Y_l$ case (solid-red line), along with the curve for a fixed entropy per baryon version of the NS LCN case (dashed-blue line) for comparison. 

To address the presence of neutrinos, it is advantageous to define a modified chemical potential $\tilde{\mu} = \mu_B + Y_l \ \mu_l$, that depends on the electronic lepton chemical potential and is equal to the Gibbs free energy per baryon of the system. This quantity has been derived and discussed in detail in Ref.~\cite{Roark:2018uls} and is shown on the horizontal axis of Fig.~\ref{Tvsmu_Bhat}. For the NS case,  $\tilde{\mu} = \mu_B$. For the PNS case, $\tilde{\mu} > \mu_B$, resulting in a lower temperature for the same modified chemical potential. The reason the temperature discontinuity is smaller in the PNS case is directly related to the smaller discontinuities between the electron and lepton chemical potentials between phases (see Fig.~3 in Ref.~\cite{Roark:2018uls} for more details). This results in a small jump in temperature across the phase transition, which violates the condition of thermal equilibrium. As explained in section 3D of Ref. \cite{Hempel:2009vp}, this is not the correct treatment and a phase coexistence region should be constructed.

\begin{figure}
\vspace{3mm}
\includegraphics[trim={1cm 0 0 2.1cm},width=9.5cm]{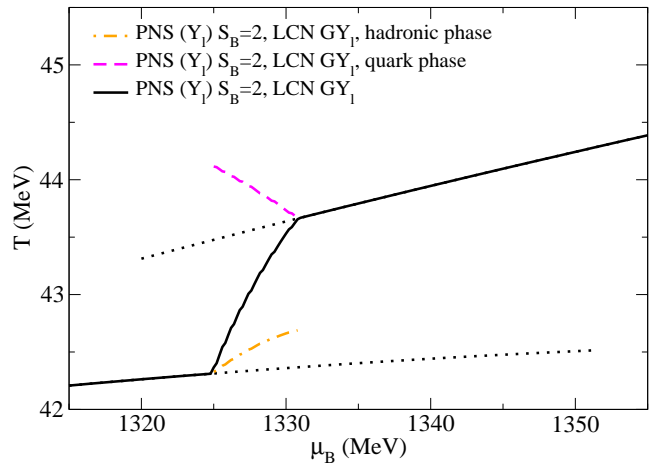}
\caption{\label{Tvsmu_B} Temperature vs baryon chemical potential phase diagram featuring PNS matter with locally conserved electric charge and globally conserved lepton fraction calculated with fixed entropy per baryon, as well as the data corresponding to the individual hadronic and quark phases in the CMF model.}
\end{figure}

Fig.~\ref{Tvsmu_B} shows similar results but focusing on the PNS LCN G$Y_l$ case. Within the $\mu_B$-range of the mixture of phases (1325.0 -- 1330.8 MeV), the results for the individual hadronic and quark phases in which case  lepton fraction is not fixed  (although entropy per baryon \textit{is} fixed and charge neutrality is satisfied) are also featured. The mixture of phases is constructed by calculating the volume of the quark phase  $\lambda$ so that global lepton fraction is conserved. Note that such a treatment still has two coexisting phases with different temperatures, not meeting the requirement of thermal equilibrium. Nevertheless, for PNS matter this jump is of the order of $1$ MeV, and a more refined approach, while conceptually gratifying, would not produce any practical improvements in the results. The temperature in the mixed phase is given by  $T_{mix} = \lambda T_Q + (1 - \lambda) T_H = 0$. The curve for the PNS LCN L$Y_l$ case is indicated by the two dotted-black lines, between which there would be a discontinuity, being that no mixture of phases is possible in this scenario. Large portions of these dotted lines are hidden from view by the black line corresponding to PNS LCN G$Y_l$. 

\begin{figure}
\vspace{3mm}
\includegraphics[trim={1cm 0 0 2.1cm},width=9.5cm]{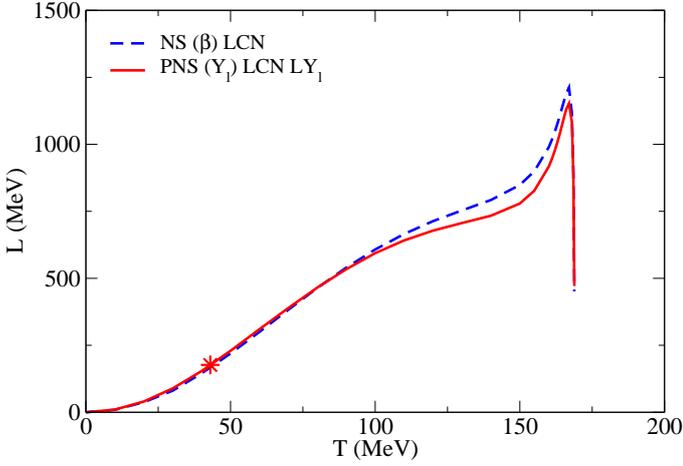}
\caption{\label{latentheat} Latent heat associated with the deconfinement phase transition vs temperature, featuring the cases of NS matter with locally conserved electric charge and PNS matter with locally conserved electric charge and lepton fraction in the CMF model. The red star around $T=$40 MeV indicates the point in the phase diagram where the deconfinement line intersects with the line for PNS matter with fixed entropy per baryon $S_B=2$.}
\end{figure}

In Fig.~\ref{latentheat}, the latent heat $L = T ~\Delta S_B$ associated with the entropy per baryon in the deconfinement phase transition is plotted over a range of temperatures $T$ for NS and PNS matter (not fixing the entropy per baryon) when both charge neutrality and lepton fraction are conserved locally. The red star in the figure indicates the point in the phase diagram where the deconfinement line intersects with the line for PNS matter with fixed entropy per baryon $S_B=2$. The curves end when their critical points are reached. Close to these points, the first order phase transition becomes weaker and latent heat decreases. The non-obvious behavior of these curves is related to the fact that the entropy density has non-trivial contributions from the order parameter of deconfinement $\Phi(\mu_B,T)$.

\begin{figure}
\vspace{3mm}
\includegraphics[trim={1cm 0 0 2.1cm},width=9.5cm]{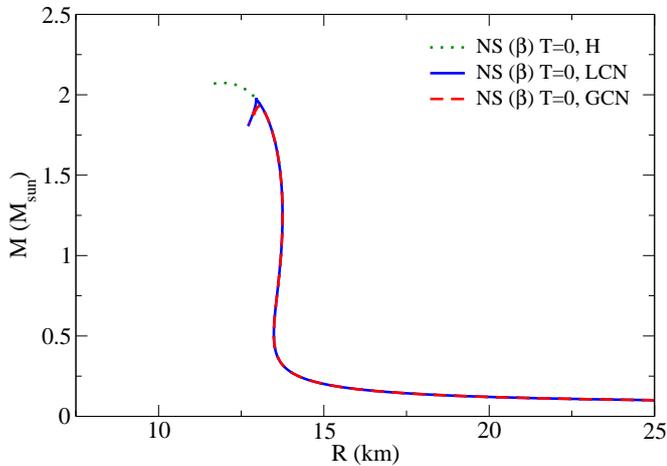}
\caption{\label{MvsR_NS+cc_T_0} Mass-Radius diagram for 3 families of stars obtained from the EoS for neutron-star matter with only hadrons, locally conserved electric charge, and globally conserved electric charge in the CMF model.}
\end{figure}

\begin{figure}
\vspace{3mm}
\includegraphics[trim={1cm 0 0 2.1cm},width=9.5cm]{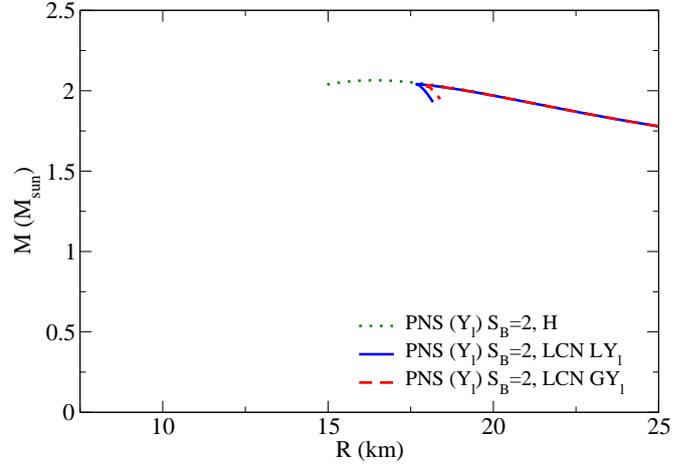}
\caption{\label{MvsR_SN+hc_fixedent} Mass-Radius diagram for 3 families of stars obtained from EoS for proto-neutron-star matter with only hadrons, locally conserved electric charge and lepton fraction, and locally conserved electric charge and globally conserved lepton fraction in the CMF model.}
\end{figure}

Next, the TOV equations are solved for each of the CMF model cases. Fig.~\ref{MvsR_NS+cc_T_0} shows the results for the three cases investigated involving NS matter, while Fig.~\ref{MvsR_SN+hc_fixedent} shows the results for the three cases involving PNS matter. As before, different crusts were used for NS and PNS's. The maximum stellar mass for each of the cases shown in Figs.~\ref{MvsR_NS+cc_T_0} and \ref{MvsR_SN+hc_fixedent} is listed in Table~\ref{maxmass}, along with the corresponding stellar radius, central (modified) baryon chemical potential, and radius occupied by hyperons, quarks, and the quark phase in the maximum-mass star. In the cases involving only hadrons (where quark matter was artificially suppressed, NS H and PNS H), the curves reach larger stellar mass values than in the cases with quarks but about the same for NS's and PNS's. In the left portion of Table~\ref{family}, the minimum (modified) central baryon chemical potential for a star to harbor hyperons, quarks, and the quark phase is listed for each case. 

From these, it becomes clear that the appearance of a quark phase renders NS's unstable, as the maximum-mass central (modified) chemical potentials are approximately equal to the thresholds for the quark phase in the NS LCN and PNS LCN LY$_l$ cases. Nevertheless, note that in the PNS LCN LY$_l$ case, quarks appear significantly before the quark phase. This is because the CMF model allows for the existence of soluted quarks in the hadronic phase and soluted hadrons in the quark phase at finite temperature. Regardless, quarks will always be the dominant component in the quark phase, and hadrons in the hadronic phase and the phases can be distinguished from one another through their order parameter $\Phi$. We believe that this inter-penetration of quarks and hadrons (that increases with temperature) is indeed physical, and required to achieve the crossover transition known to take place at small chemical potential values \cite{Aoki:2006we}. In the cases with global conserved quantities, both NS GCN and PNS LCN GY$_l$ allow stars with a couple of kms of mixtures of phases.

Hyperons ($\Lambda$ in NS's and $\Lambda$ and $\Sigma$ in PNS's) are present in some of the NS's and all of the PNS's produced, as thermal effects become more important than baryon mass differences in the latter case. $\Xi$'s do not appear in any type of star, although included in the CMF model. The presence of hyperons (quarks) can be seen when comparing the maximum-mass stellar radius in the third column in Table I and the fifth column (sixth-seventh columns) of Table~\ref{family}, with the maximum radius of stars in a family to contain hyperons (quarks). 

\begin{table*}
\begin{center}
\caption{\label{maxmass}\textbf{Single Star:} The maximum stellar mass for each case considered in the CMF model, along with the corresponding stellar radius, central (modified) baryon chemical potential, and radius inside the maximum-mass star occupied by hyperons, quarks, and a quark phase.}
\begin{tabular}[b]{c|c|c|c|c|c|c}
\hline
\hline
Case & \ \ M$_{max}$ (M$_\odot$) \ \ & \ \ corr. R (km) \ \ & \ \ corr. central $\mu_B$ (MeV) \ \ & \multicolumn{3}{c}{ \ \  corr. stellar R (km) occupied by: \ \ } \\ &&&& \ \ \ \ \ \ hyperons \ \ \ \ \ & \ \ quarks \ \ & \ \ quark phase \ \ \\
\hline
NS H & 2.07 & 11.94 & 1599.0 & 5.53 & N/A & N/A \\
NS LCN & 1.97 & 12.95 & 1345.0 & 5.53 & 0.42 & 0.42 \\
NS GCN & 1.93 & 13.07 & 1319.0 & 4.95 & 2.03 & 2.03 \\
PNS H & 2.06 & 16.38 &  1460.8 & 10.47 & N/A & N/A \\
PNS LCN L$Y_l$ & 2.04 & 17.69 & 1452.0 ($\tilde{\mu}$) & 10.47 & 7.52 & 0.44 \\
PNS LCN G$Y_l$ & 2.03 & 18.04 & 1326.0 & 10.51 & 7.42 & 0.87 \\
\hline
\hline
\end{tabular}
\end{center}
\end{table*}

\begin{table*}
\begin{center}
\caption{\label{family}
\textbf{Family of Stars:} Minimum central (modified) baryon chemical potential as well as the maximum stellar radius necessary for a star to harbor hyperons, quarks, and a quark phase in the CMF model. In the cases where electric charge or lepton fraction are conserved globally, the threshold values of central baryon chemical potentials (as well as the end range of maximum stellar radii) corresponding to the mixture of phases is instead listed in the "quark phase" columns.}
\begin{tabular}[b]{c|c|c|c|c|c|c}
\hline
\hline
Case & \multicolumn{3}{c|}{ \ \ minimum central $\mu_B$ (MeV) required for: \ \ } & \multicolumn{3}{c}{ \ \ maximum stellar R (km) required for: \ \ } \\&  \ \ \ \ \ $\Lambda$  \ \ \ \ \ & \ \ quarks \ \ & \ \ quark phase \ \ & \ \ \ \ \ hyperons \ \ \ \ \ & \ \ quarks \ \ & \ \ quark phase \ \ \\
\hline
NS H & 1229.5 & N/A & N/A & 13.42 & N/A & N/A \\
NS LCN & 1229.5 & 1345.5 & 1345.5 & 13.42 & 12.95 & 12.95 \\
NS GCN & 1229.5 & 1302.0 & 1302.0 & 13.42 & 13.13 & 13.13  \\
PNS H & 955.7 & N/A & N/A & all & N/A & N/A \\
PNS LCN L$Y_l$ & 955.7 & 1090.5 & 1452.5 ($\tilde{\mu}$) & all & 26.25 & 17.69 \\
PNS LCN G$Y_l$ & 955.7 & 1090.5 & 1325.0 & all & 26.25 & 17.69  \\
\hline
\hline
\end{tabular}
\end{center}
\end{table*}

While for NS's the appearance of a quark phase (no matter NS LCN or NS GCN) suppresses hyperons, this does not happen as strongly for PNS's. This behavior can be seen in the following figures: Fig.~\ref{NSfull_pop} features the particle population curves (as functions of the stellar radius) in the maximum-mass star obtained via the NS GCN EoS, while Fig.~\ref{SNfull_pop} features the curves in the maximum-mass star obtained via the PNS LCN G$Y_l$ EoS (values shown in the right portion of Table I). Note that the quark population curves all "kink" at one particular $R$-value for each case: 2.03 km in Fig.~\ref{NSfull_pop} and 0.87 km in Fig.~\ref{SNfull_pop} showing the beginning of the mixed phase, although down quarks persist well beyond this radius for PNS's.

Finally, we explore how changes in spinning frequency can modify the stellar particle population as stars age. We
follow the formalism proposed in \cite{Glendenning:1993di}, which includes monopole and quadrupole corrections to the metric due to rotation. We then solve self-consistently the equation for the Kepler frequency at fixed baryon number including the dragging of reference frames. Fig.~\ref{densityVSomega}  shows the stellar central baryon number density for different rotational frequencies in the CMF model. In each case [considering NS's (solid-blue line) and PNS's (dashed-red line) both with mixtures of phases], we fix the number of baryons to the respective non-rotating maximum masses. We find that, even when rotating at the maximum allowed Kepler frequency (vertical thin red line), the PNS always contains hyperons and quarks. But only when its rotational frequency goes below $89$ Hz, it develops a mixed phase, after which the fraction of the quark phase rapidly increases. On the other hand, the cold deleptonized NS rotating at the maximum Kepler frequency  (vertical thin blue line) contains only nucleons. It develops $\Lambda$-hyperons when its rotational frequency drops below $903$ Hz, and when it goes below $295$ Hz, it develops a mixed phase with quarks. 

Note that the maximum-mass non-rotating PNS in Fig.~\ref{densityVSomega} has a larger baryon number than the maximum-mass non-rotating NS. When the rotating PNS baryon number is fixed to that of the non-rotating NS, the mixed phase disappears (dotted-orange line).  Thus, an evolutionary path from a rotating PNS \textit{without} a mixed phase  (but with hyperons and quarks) to a cold-deleptonized NS \textit{with} a mixed phase is possible in the CMF model. On the other hand, a PNS with a mixed phase cannot have a NS counterpart at the same baryon number and will, therefore, collapse to a black hole.

\begin{figure}
\vspace{3mm}
\includegraphics[trim={1cm 0 0 2.1cm},width=9.5cm]{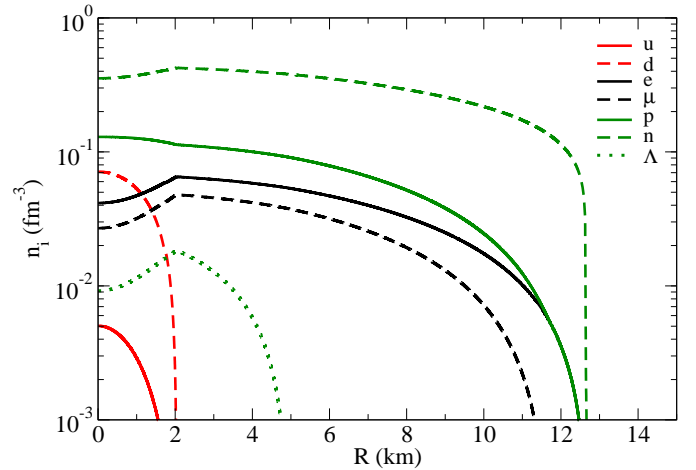}
\caption{\label{NSfull_pop} Particle number densities (quark number densities are divided by $3$) as a function of stellar radius of  the maximum-mass neutron star with globally conserved electric charge in the CMF model.}
\end{figure}

\begin{figure}
\vspace{3mm}
\includegraphics[trim={1cm 0 0 2.1cm},width=9.5cm]{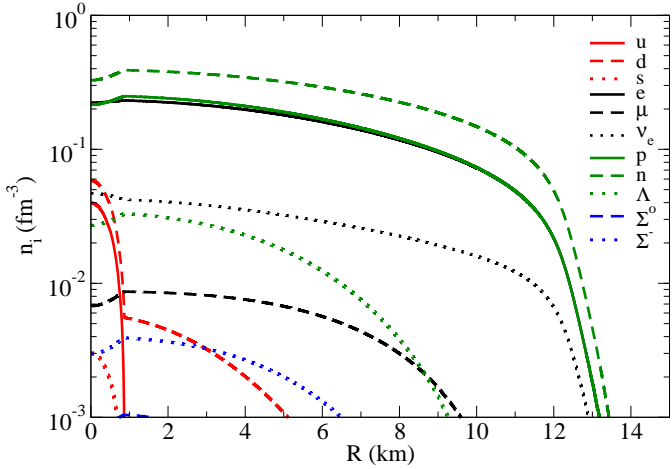}
\caption{\label{SNfull_pop} Particle number densities (quark number densities are divided by $3$) as a function of stellar radius of the maximum-mass proto-neutron star with locally conserved electric charge and globally conserved lepton fraction in the CMF model. Note that the curve for protons largely overlaps the curve for electrons.}
\end{figure}

\section{Conclusions and Outlook}

In this work, we studied protoneutron-star properties, focusing on the effects caused by different particle populations. For this purpose, we presented different models with different degrees of freedom and different features. The DSH formalism with nucleons provided several different equations of state, being presented here for the first time at fixed lepton fraction. The CMF formalism with nucleons, hyperons, and quarks generated several different equations of state, assuming different conditions for the deconfinement phase transition. For the first time results were presented for mixtures of phases inside stars with global lepton conservation and fixing the entropy per baryon in the CMF (or any other) model.

We found that, as the stellar composition becomes more complex, so do the effects of finite temperature/entropy per baryon and fixed lepton fraction on stellar properties. While in the DSH model maximum-mass values are directly related to the isospin-symmetry of nucleons, in the CMF model it is also related to the suppression/enhancement of hyperon and quark content, together with first order phase transition effects. In the CMF model, the appearance of quarks suppresses hyperons and the existence of a mixed phase allows for NS's with quarks to be stable (with up to about $2$ km of mixed phase). In PNS's, quarks are present in stable stars regardless of the existence of mixed phases and can occupy large portions of stars (up to about $7$ km).

\begin{figure}
\vspace{3mm}
\includegraphics[trim={1cm 0 0 2.1cm},width=9.5cm]{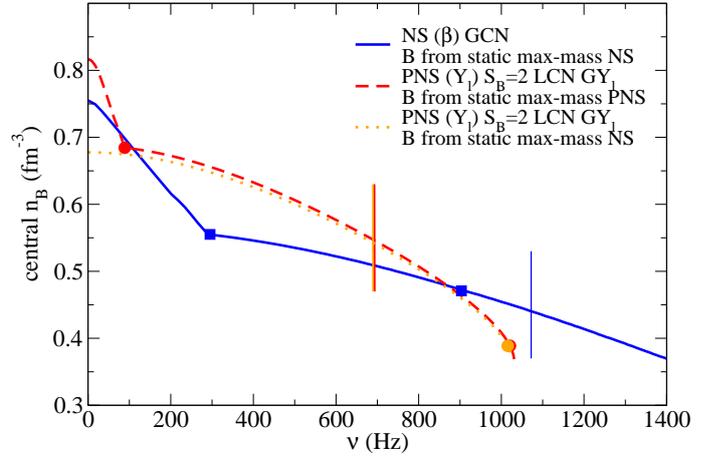}
\caption{\label{densityVSomega} Stellar central baryon number density for different rotational frequencies in the CMF model.  We fix in each curve the number of baryons to the non-rotating maximum-mass one in different evolution stages. The thin vertical lines show the Kepler limiting frequency for each case. The blue squares mark the appearance of Lambdas and quarks/quark phase in NS's, while the red/orange circles mark the appearance of quarks and quark phase in PNS's. Hyperons are always present in PNS's.}
\end{figure}

In addition, other factors have to be accounted for when studying stellar evolution, including constant number of baryons and effects due to stellar rotation. Assuming that stars do not speed up significantly as they loose neutrinos and cool down,  the CMF model predicts that large-mass stars contain a small amount of quarks (not in a mixed phase) and more hyperons in the earliest stage of their evolution and can develop a mixed phase with fewer hyperons at a later time. 

Summarizing, our results point out to the delicate and complex balance between the existence of hyperons and quarks in hot dense matter, which should play an important role in core-collapse supernovae modeling.  So far, there have been a few supernova simulations including hyperons/quarks \cite{Drago:1997tn,Sagert:2008ka,Fischer:2010wp,Nakazato:2010ue,Ouyed:2010td,Banik:2014rga,Yasutake:2012dw,Benvenuto:2013vqa,Fischer:2017lag,epsztein00,ishizuka08,baumgarte96,char15}.  However, to our knowledge, consistent simulations of neutron stars from birth in the core-collapse, through the PNS stage and cooling to cold NS's using the same model have not been reported as yet. It is our goal to perform such a study utilizing the EoS models  presented in this work. Moreover, we plan to supplement the results presented here by studying the EoS based on the Quark-Coupling model  \cite{Guichon:2018uew}, which includes the full baryon octet. Finally, a natural more simple extension of our work is to study magnetic field effects on the stellar particle population. Work along these lines is already under way.

\section*{Aknowledgements}

VD, CC, and JR  are supported by the National Science Foundation under grant PHY-1748621. XD and AWS were supported by DOE SciDAC grant DE-SC0018232 and NSF grant PHY 1554876.

\bibliographystyle{mnras}
\bibliography{paper}
\end{document}